# Human–AI Collaboration in Science at Scale: A Global Large-scale Randomized Field Experiment

Binglu Wang[1,2,3*], Weixin Liang[4], Jiahui Xue[1], Yuhui Zhang[4], Hancheng Cao[5], Dashun Wang[1,2,3*], Yian Yin[6*]

[1] Kellogg School of Management, Northwestern University, Evanston, IL, USA

[2] Center for Science of Science and Innovation, Northwestern University, Evanston, IL, USA

[3] Northwestern Institute on Complex Systems, Northwestern University, Evanston, IL, USA

[4] Department of Computer Science, Stanford University, Stanford, CA, USA

[5] Goizueta Business School, Emory University, Atlanta, GA, USA

[6] Department of Information Science, Cornell University, Ithaca, NY, USA

* Corresponding author. Email: binglu.wang@kellogg.northwestern.edu, dashun.wang@kellogg.northwestern.edu, yian.yin@cornell.edu

**Collaboration is the defining mode of modern science, yet its core mechanism—feedback—remains hard to observe, difficult to scale, and unequally distributed. Here we test whether large language models (LLMs) can contribute to this hidden but vital practice and reallocate scientific feedback, an essential yet scarce resource for knowledge production. In a global large-scale randomized field experiment, we delivered customized LLM-generated feedback for over 31,000 arXiv preprints across 150 fields and more than 45,000 researchers from 133 geographic regions. Relative to controls, authors who received feedback had a significantly higher likelihood of revising their manuscripts, corresponding to a 12.55% relative increase over the baseline revision rate. Exposure to AI feedback also increased authors' subsequent use of LLM tools in their future papers, suggesting longer-run shifts in scientific practice. These effects were strongest among authors from non-English-dominant research regions, manuscripts less embedded in the scholarly literature, and teams with lower *h*-indexes and earlier career stages, consistent with the idea that AI feedback may provide the greatest benefit where access to timely critique is otherwise limited. Together, these findings provide causal evidence that structured AI-based interventions can transform access to scientific feedback from a largely private advantage into a more widely distributed resource, with broader implications for productivity, equity, and capacity across the global research system.**

Collaboration has become the defining mode of modern science. Across nearly every field, teams drive the majority of discoveries, and collaborative work underpins many of the most important innovations and breakthroughs [1-4]. This growing importance also reflects the changing reality of science, as scientific challenges expand in scale and complexity [5, 6] while individual scientists increasingly narrow their focus and expertise [7] or face difficulties in branching into new terrains [8]. By combining perspectives and approaches, collaboration enables scientists to generate ideas, refine methods, and strengthen conclusions, making it indispensable for progress.

At the heart of collaboration lies feedback, through which scientists challenge assumptions, surface anomalies, and sharpen contributions [9-11]. Science often advances when ideas bounce off each other, yet despite its central role, this essential collaborative practice is largely invisible in data, taking place through various formal and informal exchange channels that rarely leave a trace. Its hidden nature makes feedback difficult to observe, to scale, and to intervene in, despite its influence on the direction and quality of scientific work.

Further, access to feedback has been deeply unequal. Opportunities to exchange ideas are stratified along geographic, linguistic, and institutional lines [9, 12, 13]. Elite networks often provide dense, continuous flows of critique, while early-career scholars outside English-dominant or resource-rich settings face systemic barriers to receiving timely and constructive feedback [14-16]. Feedback thus drives progress but also reproduces inequality in science, shaping whose ideas are refined and amplified, and whose remain underdeveloped.

Against this backdrop, large language models (LLMs) are increasingly capable of synthesizing information and offering critique [17-20]. This raises a provocative question: Can AI participate in this hidden but vital collaboration, and reallocate scientific feedback, an essential resource for knowledge production, at global scale? If LLMs can provide useful feedback, they might expand collaborative capacity and reduce disparities in who can benefit from it. Yet at the same time, there is substantial skepticism about whether they can play a meaningful role in science. While LLMs have demonstrated measurable gains in codified domains such as coding, writing, and professional decision-making [21-26], conducting scientific research differs in fundamental ways. It is tacit, relying on context-sensitive judgments and accumulated expertise; it is highly specialized, with each field defining its own standards of rigor and contribution; and it is credibility-dependent, where the value of feedback hinges on whether it is trusted and defensible to expert peers. And ultimately, scientific research is a venture into the unknown. These features make collaboration with AI far less straightforward. Critics argue that AI feedback is often too generic, not credible, or prone to errors, raising doubts about whether it could meaningfully influence scientific practice [27, 28]. Others worry that reliance on AI could dilute the human expertise on which discovery depends [29, 30].

Here we shed light on this question through a global large-scale randomized field experiment on more than 31,000 scientific manuscripts. Conducted in early 2024, our intervention occurred during the brief onset phase of LLM adoption in scientific writing [31, 32], after the tools became widely available but before they were routinely embedded in research workflows. This moment allows us to capture a clean view of how AI enters a core collaborative practice. We exogenously assign papers to receive a standardized bundle of customized LLM-generated feedback produced under a fixed protocol, allowing us to isolate the causal effect of receiving AI feedback rather than the effects of self-selected tool adoption or differences in prompting skill. By delivering the feedback to authors across 150 scientific fields and 133 geographic regions, we evaluate whether AI can meaningfully contribute to the collaborative practice of offering feedback. We examine both short-term effects on revisions and longer-term effects on adoption, with special attention to disparities across linguistic, institutional, and career contexts. In doing so, we provide causal evidence on whether and how structured AI-based interventions can engage in the collaborative practices that underpin scientific progress.

## Experiment Design

We examine a collaboration setting that mirrors how researchers draft manuscripts and seek critique from coauthors. This process plays a central role in advancing research, and allows us to observe how collaboration shapes subsequent outcomes. To operationalize this process, we introduced an institutionalized "AI collaborator" at scale, delivering structured yet customized feedback on scientific drafts. Recent advances have shown that LLM-based systems can generate comments resembling those of peer reviewers, with quality validated by experts and even preferred to some human reviews [17, 33]. Building on these advances, our study moves beyond validation and reviewer assistance to evaluate whether AI-generated feedback can influence scientific production itself, delivering it directly to authors while their manuscripts are still in development[1].

[1] The study was reviewed and approved by the Institutional Review Board at Northwestern University (STU00220102.)

**Sample collection and randomization:** We collected preprints initially posted on arXiv between January and June 2024, and randomized at the paper level, stratified by 150 fields (Fig. 1). The initial randomized cohort comprised 34,340 manuscripts posted as first versions, from which we extracted 50,454 authors with valid email addresses across 133 regions worldwide (SM Fig. S1). During the period between feedback generation and delivery, 3,320 manuscripts (9.7%) posted an updated version. Because feedback was generated based on each paper's initially posted version, it could not plausibly inform revisions that had already occurred. We thus focus our analyses on preprints that had only the first version when feedback was delivered, to align our analysis with our causal estimand (the effect of receiving feedback for initial manuscripts on subsequent revision). The overall analysis sample contains 31,020 manuscripts (treatment group: $N = 15{,}542$; control group: $N = 15{,}478$) and 45,466 contacted authors. Baseline covariates remain balanced after restriction (SM Table S1). This risk-set definition follows standard causal-inference practice, aligning time-zero eligibility with treatment assignment and ensuring that only manuscripts genuinely at risk of exposure contribute to the estimated effect.

**Treatment and control conditions:** Authors in the treatment group received an email (see email template in SM Section S2) containing a private link to an interactive webpage with customized AI-generated feedback on their preprint produced under a fixed protocol, while control authors received no intervention. Feedback was generated using a state-of-the-art LLM framework, augmented with a multi-agent pipeline that (i) conducted review analysis, (ii) suggested alternative titles, and (iii) checked grammar. We further apply in-context learning to tailor the comments toward the specific content of each preprint (see feedback examples in SM Section S1.3 and prompt detail in SM Section S10.1). For security and privacy, feedback links were encrypted, and access was restricted to the corresponding author via a private webpage.

**Timeline:** To manage the logistics of generating and distributing feedback at scale, we divided the sample into two cohorts. The first cohort included preprints first posted on arXiv between January and March 2024 ($N = 14{,}905$); feedback, generated on each paper's first posted version, was delivered in June 2024. The second cohort consisted of preprints first posted between April and June 2024 ($N = 16{,}115$), with feedback delivered in August 2024. Authors therefore received feedback between one and five months after initial posting.

**Outcome measures:** We trace two major outcome measures. (1) Short-term revisions: The number of revised versions posted within a month of feedback delivery. Revision activity is measured using the version history records available on the arXiv platform for each manuscript. (2) Long-term adoption: Whether authors subsequently adopted LLM tools in their own writing within a year of feedback delivery. To assess this, we collected all publications posted by treatment and control authors in a twelve-month window after the intervention and applied a state-of-the-art AI detection model [31] to detect LLM-generated content. Detailed descriptions of these measurements are provided in SM Section S4.

Together, this design enables us to test whether AI can meaningfully reshape the collaborative infrastructure of feedback, influence revisions in the short term, and shape the adoption of new tools in the longer term.

**Results**

**AI feedback increases short-term revision activity.** We first test whether feedback matters by examining whether it alters the scientific work. In this respect, revisions offer a clear, observable signal of engagement because they require deliberate effort and a decision to act on critique. To this end, we assess revision behaviors by tracking the number of manuscript revisions that were updated on arXiv following the intervention. Using time-stamped records, we measure the number of revised versions authors submitted within a one-month window after receiving feedback, and fit an OLS regression model:

$$Revision_i = \beta_0 + \beta_1 \times Treatment_i + \varepsilon_i,$$

where $Revision_i$ is the number of revised versions paper $i$ has post-treatment and $Treatment_i$ denotes assignment to the AI-feedback condition. The coefficient $\beta_1$ captures the intent-to-treat effect of receiving AI-generated feedback on revision behaviors. Receiving AI-generated feedback significantly increased short-term revision activity (Fig. 2A, $p < 0.05$, SM Table S2). Treated papers exhibited an average increase of 0.005 revisions within one month relative to the control group ($p < 0.05$), corresponding to a 12.55% increase over the baseline revision rate. Although the absolute magnitude appears small, reflecting the low base rate of short-term revisions, the relative increase indicates that AI feedback meaningfully altered author behavior at scale. Estimates remain consistently positive across alternative observation windows (SM Fig. S2).

To examine what authors revised, we developed a computational pipeline that compares original and revised arXiv manuscripts and classifies edits into substantive content updates versus language, typographical, and formatting changes (SM Section S3). Our exploratory analysis shows that, differences in surface-level edits are minimal (SM Fig. S6A; SM Table S5). If any, treated authors exhibit a higher frequency of substantive revisions, indicating directional increases in conceptual changes such as ethics and novelty (SM Fig. S6B; SM Table S6).

**Subgroup patterns across institutional, manuscript and author characteristics.** The aggregate treatment effect raises a natural question: who benefits most from AI-generated feedback? If AI functions as a scalable source of critique, its impact may vary depending on authors' access to existing feedback infrastructures and their position within the scientific system. To examine this possibility, we estimate treatment effects across four dimensions capturing institutional environment, manuscript positioning, and author experience: language environment, scholarly embeddedness, career age, and prior prominence ($h$-index).

These moderators reflect different channels through which access to feedback may vary. Institutional language environment influences participation in global scientific communication [32, 34, 35]. Manuscript-level scholarly embeddedness, measured by number of references cited in the initial version, captures how extensively a paper is situated within existing scholarly conversations, and may shape its responsiveness to external critique. Author-level characteristics, including career age and $h$-index, approximate accumulated experience and prominence within the scientific community (SM Section S4 for method details).

Across all dimensions, treatment effects follow a consistent pattern in which statistically significant estimates appear in the lower category of each moderator (Fig. 2B; SM Table S3). The institutional language environment is captured by English-dominant status, defined as whether at least one contacted author is affiliated with a country whose official language includes English. Among manuscripts affiliated exclusively with non–English-dominant environments, AI feedback increases revisions by 0.007 ($P < 0.05$) corresponding to a 19.9% relative increase over control baseline. A comparable effect is observed among manuscripts with lower embeddedness (0.007; $P < 0.05$), corresponding to a 26.4% increase relative to the control group. At the author level, statistically significant effects are likewise concentrated among authors with lower $h$-index (20.2% increase; $P < 0.05$) and earlier career age (18.9% increase; $P < 0.05$).

In contrast, the corresponding estimates for manuscripts and authors in the higher category (i.e., English-dominant environments, higher scholarly embeddedness, higher $h$-index, and longer career age) are small and statistically indistinguishable from zero ( $P =$ $0.98, 0.63, 0.31,$ and $0.28$). Robustness checks that alter the post-treatment window (one to two months, SM Section S5; SM Fig. S3-S4) yield largely consistent results, with authors consistently showing revision effects and the effects are consistently larger for low-category groups.

Across all four moderators, statistically significant estimates appear primarily among manuscripts and authors in the lower category of each moderator, whereas estimates for higher category groups are small and not statistically distinguishable from zero. Nevertheless, the consistent patterns across institutional, manuscript, and author characteristics suggest that the aggregate revision response is largely driven by manuscripts and authors situated further from established linguistic, intellectual, or professional advantage.

**AI feedback and subsequent adoption of LLM tools.** Although our intervention occurred only once, its influence may extend beyond immediate behavior [36]. This raises an interesting question of whether exposure to AI-generated feedback leaves detectable traces in later research practice. To this end, we examined whether exposure to AI-generated feedback increased subsequent adoption of LLMs. We tracked each author's arXiv preprints within a twelve-month window after the intervention and analyzed abstracts with a widely used AI detection model [31], which assigns a score $\alpha$ estimating the fraction of text likely generated by LLMs. To isolate new uptake, we restrict the analysis to authors with minimal prior use ($\alpha_{pre} \leq \tau$, where $\tau = 0.1$ in our primary analysis). We estimated the treatment effect using the same specification at the author level:

$$Adoption_p = \beta_0 + \beta_1 \times Treatment_p + \varepsilon_p,$$

where $Adoption_p$ is the average $\alpha$ of post-treatment papers of researcher $p$. $Treatment_p$ denotes assignment to the AI-feedback condition and the coefficient $\beta_1$ captures the intent-to-treat effect of receiving AI-generated feedback on the likelihood of later LLM adoption.

We find that, among authors whose pre-treatment writing showed minimal detectable LLM involvement, treated authors exhibit significantly higher subsequent LLM adoption than controls ($\alpha_{treatment} = 0.150$, $\alpha_{control} = 0.142$; Fig. 3A), corresponding to a 5.29% relative increase in adoption intensity within twelve months ( $P < 0.05$; SM Table S7).

Heterogeneity patterns closely parallel those observed for revisions (Fig. 3B). Treatment effects are again statistically significant among authors affiliated with non–English-dominant environments (8.20% increase; $P < 0.01$), those with lower scholarly embeddedness (6.34% increase; $P < 0.05$), lower $h$-index (6.98% increase; $P < 0.05$), and earlier career age (6.16% increase; $P < 0.05$). In contrast, the corresponding estimates for English-dominant authors and those with higher embeddedness, $h$-index, or career age are attenuated and not statistically distinguishable from zero ($P = 0.610, 0.124,\ 0.325, 0.217$).

Robustness checks that alter the post-treatment window (six to twelve months) or vary the adoption threshold $\tau$ (0 to 0.2) yield broadly consistent results (SM Section S5; SM Fig. S7-S9). Across specifications, authors show substantial and statistically significant adoption effects, while estimated effects remain larger among low-category groups. These adoption patterns mirror the revision findings, suggesting that the same groups that showed increased engagement with substantive revisions are also more likely to incorporate AI into their research workflows over the longer term. The consistent pattern between revision and adoption results suggests that the observed effect does not merely reflect transient experimentation but rather sustained integration, as measured by continued use across subsequent manuscripts.

Together, these results demonstrate that AI feedback can shape collaborative practices at scale, not only prompting immediate revisions but also accelerating the diffusion of AI into research workflows. Both shorter- and longer-term effects were strongest among those historically excluded from dense feedback networks, suggesting that AI can help lower barriers to participation and act as a catalyst for broader adoption of new tools. In this way, AI functions not only as a short-term collaborator but also as an enduring influence on how scientific work is conducted.

## Discussion

This study provides causal evidence that large language models can meaningfully participate in a central collaborative practice of science: offering feedback. In a global large-scale field experiment, exposure to AI-generated critique prompted substantive revisions to manuscripts and increased subsequent adoption of AI tools. These effects were not evenly distributed. They were stronger among manuscripts and authors positioned further from established linguistic, intellectual, or professional advantage, suggesting that AI has the potential to act as a collaborative equalizer in science, expanding access to critique where traditional channels are less accessible.

These findings should be interpreted in light of inherent design trade-offs that accompany field experimentation at this scale. Our design compares treated authors to a no-contact control group, raising the question of whether our results reflect feedback content or the attentional salience of receiving any correspondence. A content-free placebo condition could in principle help disentangle the two, but mass emails without substantive content risk being dismissed as spam, undermining their validity as a counterfactual. However, the increase in long-term LLM adoption, combined with our exploratory content analysis on participant feedback and paper revision (directional increases in substantive, conceptual revisions rather than merely surface-level edits), suggests that authors meaningfully engaged with the actual content of the AI critique, making it unlikely that the observed changes are driven purely by an attention mechanism. Our adoption measure relies on LLM-detection methods [31], which are inherently noisy and may exhibit differential accuracy, such as for non-native English writing styles [35]. Since our treatment is exogenously assigned, such systematic bias should be balanced across the arms. Consequently, the observed differences between these groups effectively isolate the causal effect of the AI feedback intervention, though future work should continue to triangulate this measure with ground-truth data on actual tool use. Further, our estimates reflect intent-to-treat effects from a single light-touch intervention, necessarily including authors who did not open or engage with the feedback; they are therefore best interpreted as a conservative lower bound on the effect among active compliers.

A key contribution of our design lies in isolating the effect of AI feedback from the heterogeneous ways researchers typically engage with generative AI. The distributional consequences of LLMs remain unsettled in prior work [21, 22, 25, 26, 37-39], partly for two reasons. First, exposure to and adoption of LLMs can be endogenous, as individuals differ in access, willingness to adopt, and the necessary skills or resources to obtain meaningful AI assistance. Second, even when adoption occurs, uses of AI can be highly heterogeneous. As a general-purpose technology, the value of LLMs depends sharply on prompting strategies, iteration, editing effort, and workflow integration [40]. Together, these sources of variation make it difficult to determine whether observed productivity gains reflect the technology itself or differences in users' ability to harness its benefits.

Our study addresses both challenges by design: Conducted in early 2024, our experiment captures scientific practice at the onset of LLM adoption, when these tools were widely available, but not yet fully formalized into drafting and collaboration workflows. This produced a natural mix of researchers, with some already experimenting with LLMs but many still relying on traditional feedback channels. This provided a rare window for effective randomization, allowing us to isolate the impact of AI feedback before endogenous adoption became widespread. Further, rather than treating generative AI as a tool individuals choose how to use, we conceptualize it as a centrally deployable resource generator. By exogenously assigning manuscripts to receive a standardized bundle of AI-generated feedback under a fixed protocol, with variation driven only by manuscript content rather than by users' prompting skill, our design removes the second source of variation. In this controlled setting, we observe a consistent equalizing pattern: authors in the low-category groups disproportionately benefit. These results not only help empirically clarify the ongoing debate on AI and productivity, but also provide new methodological leverage for studying how AI reshapes inequalities in knowledge production.

Importantly, the consequences of AI feedback extended beyond immediate revisions. Treated authors not only showed suggestive evidence of reinforcing the intellectual core of their manuscripts, with directional increase in incorporating more changes addressing ethics and novelty, but also demonstrated a greater likelihood of integrating AI tools into future writing. Qualitative responses reinforced these patterns: participants described the feedback as clear and useful, and many expressed an intention to continue using such tools (SM Section S7), consistent with the hypothesis that a single exposure may not only shape the content of individual manuscripts but also leave lasting traces in research practices [41]. As technology evolves [42, 43][2], adoption accelerates and baseline use rises, the absolute magnitudes of our effects may evolve, but the underlying mechanisms we document, including expanded access to critique and subsequent tool adoption, offer a timely baseline for understanding how human-AI collaboration takes hold in the increasingly AI-infused global research ecosystem.

Our study also advances efforts to understand and experiment with the scientific process itself. Field experiments in science are, in general, difficult to execute or scale, with most prior trials limited to small samples or narrow domains [44-47]. By leveraging bibliometric infrastructure and automated feedback systems, we demonstrate the feasibility of running global large-scale, personalized field experiments on live scientific practice. This opens new opportunities to not only observe scientific behavior retrospectively but to test interventions capable of shaping how knowledge is produced, revised, and adopted across global communities.

Finally, our findings raise broader questions about the future of collaboration in an era of increasingly AI-integrated research. If AI can reliably provide critique on clarity, theory, and prior work, human collaborators may shift toward higher-order conceptual and design roles, but marginal collaborations may also be displaced. A further challenge arises from standardization: as more researchers lean on similar underlying models, science risks a form of homogenization [17], with diverse lines of critique converging toward a narrow band of recommendations. This concern parallels emerging evidence that while AI can raise individual performance, it may also reduce collective diversity unless deliberately counteracted [48, 49]. These possibilities highlight the importance of designing human-AI collaboration to preserve heterogeneity, through plural model architectures, varied objectives, and institutional norms that reward dissent and methodological diversity.

[2] Other AI feedback tools such as https://paperreview.ai/ and https://www.refine.ink/.

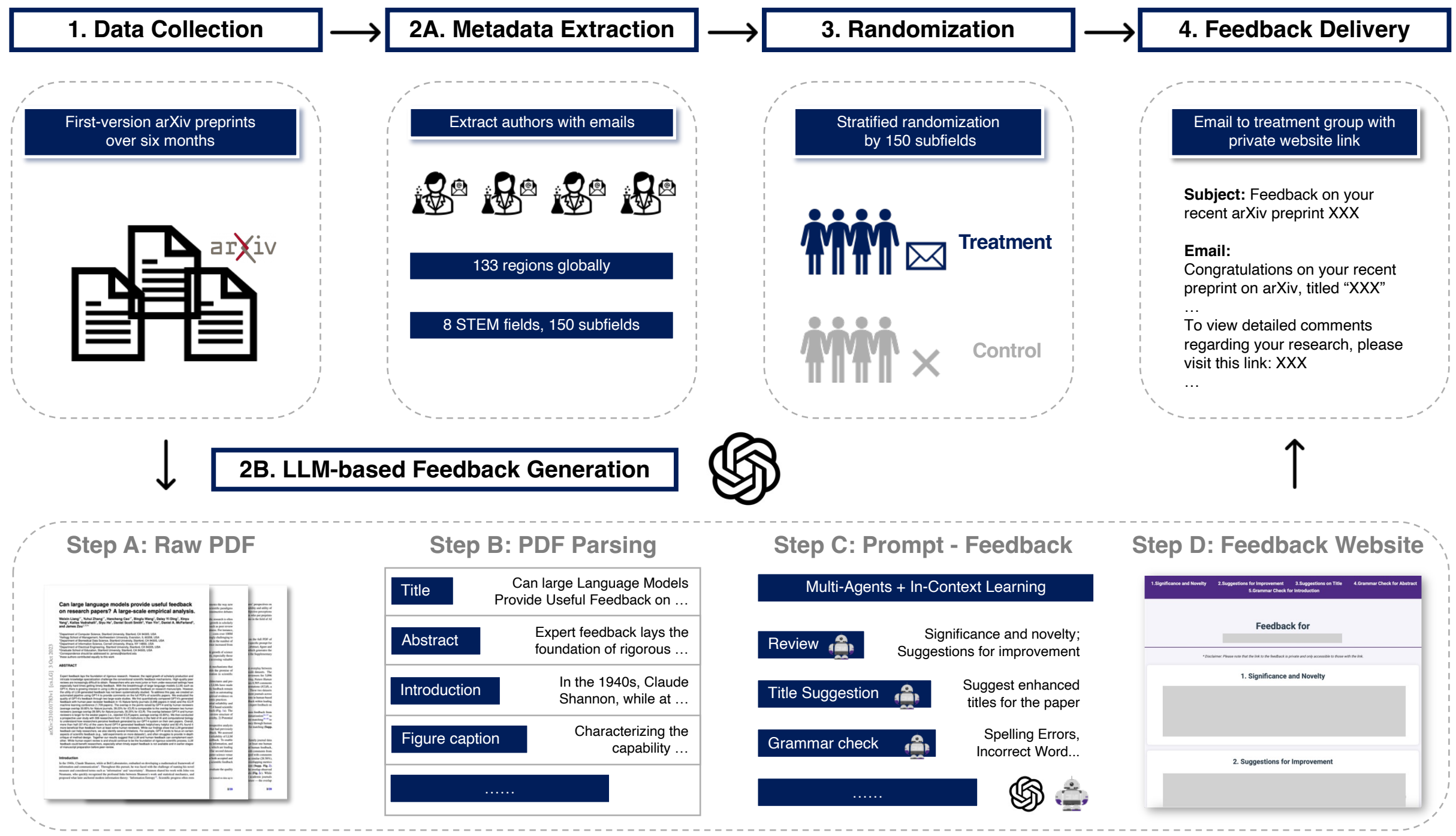

**Fig. 1. LLM feedback generation and experiment workflow. (Step 1)** We collected all recent arXiv preprints (Jan–Jun 2024) for which only the first version was available. **(Step 2)** For each preprint, we extracted metadata (authors, affiliations, field; **Step 2A**), and parsed structured text from the raw PDFs to generate customized feedback using an LLM-based system **(Step 2B)**. **(Step 3)** We implemented a stratified randomization process across 150 fields, assigning each paper to either the treatment or control group. **(Step 4)** Each author in the treatment group receives a personalized email, with a private link giving them access to the AI-generated feedback.

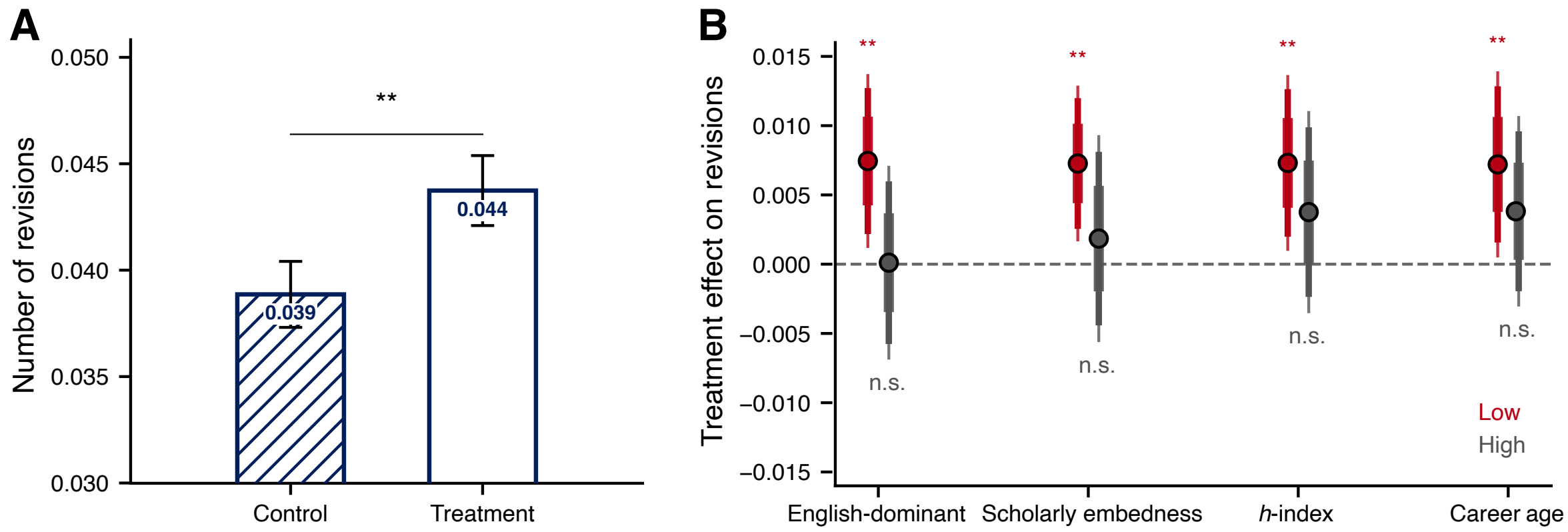

**Fig. 2. AI-generated feedback increases short-term manuscript revisions and exhibits heterogeneity. (A)** Manuscripts assigned to receive AI-generated feedback show a higher revision rate than controls (0.044 vs. 0.039), corresponding to a 12.55% relative increase over baseline. Bars represent group means and error bars denote ±1 standard deviation. **(B)** Subgroup estimates of the treatment effect across institutional and author characteristics. Points represent estimated treatment effects from regression models and vertical lines denote the 95% and 90% confidence intervals, as well as one standard deviation. Red markers correspond to the lower category of each moderator while grey markers correspond to the higher category. Treatment effects are statistically significant among manuscripts with lower scholarly embeddedness and among authors affiliated with non–English-dominant environments, with lower H-indices and shorter career ages. In contrast, the effects are not statistically distinguishable from zero among their higher-resource counterparts. Standard significance thresholds are indicated: $*** P < 0.01$, $** P < 0.05$, $* P < 0.1$.

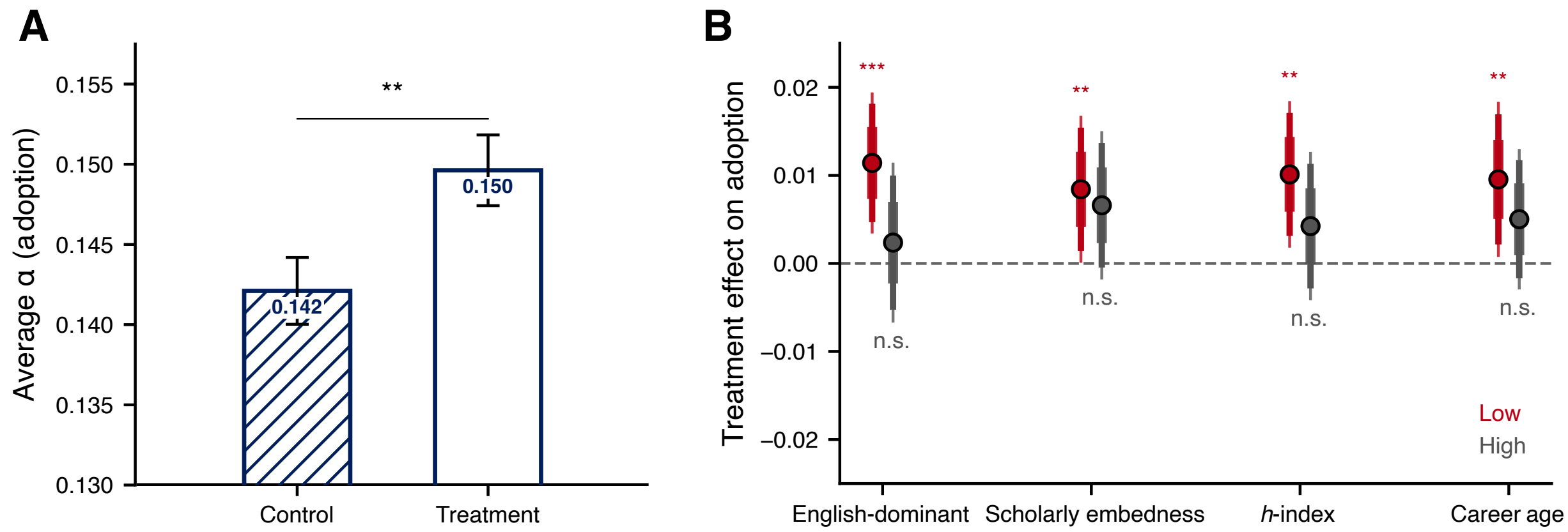

**Fig. 3. Treatment effect on LLM adoption, measured by $\alpha$. (A)** Average LLM adoption intensity ($\alpha_{post}$) within one year following feedback delivery, among authors with minimal prior LLM usage ($\alpha_{pre} \leq 0.1$). $\alpha$ measures the estimated proportion of manuscript content generated or modified by LLM tools, following Liang et al. (2025). Authors assigned to receive AI-generated feedback exhibit higher subsequent adoption than controls. Bars represent group means and error bars denote ±1 standard deviation. **(B)** Subgroup estimates of the treatment effect on $\alpha_{post}$ across author characteristics. Red markers denote the lower category of each moderator, and grey markers denote the higher category. Statistically significant positive treatment effects are observed among authors affiliated with non–English-dominant environments, those with lower average scholarly embeddedness, lower h-index, and shorter career age. In contrast, the corresponding estimates for higher-category groups are not statistically distinguishable from zero. Statistical significance thresholds: $*** P < 0.01$, $** P < 0.05$, $* P < 0.1$.